\documentclass[epj]{svjour}
\usepackage{graphics}
\begin{document}
\title{Precision calculation of energy levels for four-valent Si I}
\author{R. T. Imanbaeva\inst{1,2} \and M. G. Kozlov\inst{1,2} \and E. A. Konovalova\inst{1}
}                     
\offprints{R.T. Imanbaeva}
\mail{raykhan.phys@gmail.com}        
%
\institute{Petersburg Nuclear Physics Institute, Gatchina 188300, Russia \and
St.~Petersburg Electrotechnical University ``LETI'', Prof. Popov Str. 5,
St.~Petersburg, 197376, Russia}
\date{Received: \today / Revised version: date}
%
\abstract{We report results of the calculation of the low-lying levels of
neutral Si using a combination of the configuration interaction and many-body
perturbation theory (CI+MBPT method). We treat Si I as an atom with four
valence electrons and use two different starting approximations, namely
$V^{N-2}$ and $V^{N-4}$. We conclude that both approximations provide
comparable accuracy, on the level of 1\%.
} 
\maketitle
\section{INTRODUCTION}
\label{intro}

The method, which combines configuration interaction (CI) and many-body
perturbation theory (MBPT) was suggested in \cite{DFK96a} and described in
more details in \cite{DFK96b}. In this approach, which is known as CI+MBPT
method, the electrons in an atom are divided into core and valence electrons.
Correlations between valence electrons are treated using CI method. The
core-core and core-valence correlation are taken into account using
second-order MBPT. During the last two decades this method was used for
numerous calculations of many-electron atoms and ions and proved to be very
effective for systems with two-three valence electrons (see, e.g.
\cite{PKRD01,SJ02,PSK12,KonKoz15}), where this method typically provide an
accuracy about 1\%, or better. Less often this method was used for systems
with four, or more valence electrons
\cite{Dzu05a,Dzu05b,Ber11,SDFS14b,Sav16,PKST16} and it is still not clear what
accuracy one can expect in this case.

There are several reasons why the accuracy and effectiveness of the method may
decrease for the four valent systems. First, the size of the CI space for a
given length of the basis set grows exponentially with the number of valent
electrons. Therefore, it may be impossible to saturate CI space. Second, in
order to start with a sufficiently good initial approximation, one needs to
include (at least partly) the field of the valence electrons in the initial
potential. This leads to the whole new class of the MBPT diagrams
\cite{DFK96b} and makes results less stable. Finally, for the four-valent
atoms there are effective three-electron interactions. This makes calculations
much more difficult \cite{BFK08,KSPT16}.

In this paper we apply CI+MBPT method to Si atom, which we treat as a
four-valent system. We compare results for $V^{N-4}$ and $V^{N-2}$ initial
approximations, where $N$ is the total number of electrons. The former
approximation includes only the field of the core electrons, while the latter
one includes also the field of the two valence $3s$ electrons. We conclude
that both approximations lead to comparable accuracy about one percent. For
the two lowest singlet states the accuracy is lower, most likely, because of
the cusp effect, which is very difficult to reproduce in the CI calculations.

%

\begin{table*}
\caption{Low lying energy levels of Si I in the $V^{N-2}$ approximation
\cite{Dzu05a} \cite{PKST16}. The energies (in cm$^{-1}$) are counted from the
ground state. Results of the CI and CI+MBPT calculation are given in columns
labeled ``CI''and ``CI+MBPT''. Corresponding relative differences of these two
calculations with the experiment \cite{NIST} are given in percent. }
\label{tab:1}       
\begin{center}
\begin{tabular}{rrrrrrr}
\hline \hline
 &     &   &   &  &
 \multicolumn{2}{c}{Differences (\%)}   \\
 \multicolumn {1}{c}{Conf.}
 &\multicolumn {1}{c}{Term}
 &\multicolumn {1}{c}{Exper.}
 &\multicolumn {1}{c} {CI}
 &\multicolumn {1}{c} {CI+MBPT}
 &\multicolumn {1}{c} {CI}
 &\multicolumn {1}{c} {CI+MBPT}\\
 \hline
 3s$^{2}$3p$^{2}$ & $^{3}P_{1}$ & 77 & 77 & 76 & 0.24 & 0.84 \\
 3s$^{2}$3p$^{2}$ & $^{3}P_{2}$ & 223 & 225  & 224 & -0.60 & -0.15 \\
 3s$^{2}$3p$^{2}$ & $^{1}D_{2}$ & 6299 & 6909 & 6469 & -9.69 &-2.71 \\
 3s$^{2}$3p$^{2}$ & $^{1}S_{0}$ & 15394 & 16621 & 15931 & -7.97 & -3.49 \\
 3s3p$^{3}$ & $^{5}S^{o}_{2}$ & 33326 & 31472 & 32917 & 5.56 & 1.23 \\
 3s$^{2}$3p4s & $^{3}P^{o}_{0}$ & 39683 & 39122 & 39773 & 1.41 &-0.23 \\
 3s$^{2}$3p4s & $^{3}P^{o}_{1}$ & 39760 & 39201 & 39851 & 1.41 & -0.23 \\
 3s$^{2}$3p4s & $^{3}P^{o}_{2}$ & 39955 & 39395  & 40045 & 1.40 & -0.22\\
 3s$^{2}$3p4s & $^{1}P^{o}_{1}$ & 40992 & 40562 & 41136 & 1.05 & -0.35 \\
 3s3p$^{3}$ & $^{3}D^{o}_{1}$ & 45276 & 44276 & 45090 & 2.21 & 0.41 \\
 3s3p$^{3}$ & $^{3}D^{o}_{2}$ & 45294 & 44295 & 45108 & 2.21 & 0.41 \\
 3s3p$^{3}$ & $^{3}D^{o}_{3}$ & 45322 & 44326 & 45138 & 2.20 &  0.41 \\
 3s$^{2}$3p4p & $^{1}P_{1}$ & 47284 & 46536 & 47519 & 1.58 & -0.50 \\
 3s$^{2}$3p3d & $^{1}D^{o}_{2}$ & 47352 & 46478 & 47314 & 1.84 & 0.08 \\
 3s$^{2}$3p4p & $^{3}D_{1}$ & 48021 & 47306 & 47722 & 1.49 & 0.62 \\
 3s$^{2}$3p4p & $^{3}D_{2}$ & 48102 & 47390 & 47805 & 1.48 & 0.62 \\
 3s$^{2}$3p4p & $^{3}D_{3}$ & 48264 & 47552 & 47966 & 1.48 & 0.62 \\
 3s$^{2}$3p4p & $^{3}P_{0}$ & 49028 & 48381 & 48768 & 1.32 & 0.53 \\
3s$^{2}$3p4p & $^{3}P_{1}$ & 49061 & 48409 & 48797 & 1.33 & 0.54 \\
\hline\hline
\end{tabular}
\vspace*{5cm}  
\end{center}
\end{table*}

\section{METHOD}
The CI+MBPT method was realized in a number of computer codes (e.g.
\cite{SJ02,Dzu05a,Ber11}). The version we use here is based on the
Hartree-Fock-Dirac (HFD) code \cite{BDT77} and the CI code \cite{KT87}. The
whole package was recently published in Ref.\ \cite{KPST15}.

In the CI method, the many-electron wave function with a given total angular
momentum $J$ is obtained as a linear combination of the many-electron Slater
determinants:
 \begin{equation}\label{eq1}
 \Psi_{J}=\sum_{i}c_{i}\Phi_{i}\,,
 \end{equation}
where determinants $\Phi_{i}$ are formed from the valence basis orbitals. In
our case these orbitals are calculated by solving HFD equations either in the
$V^{N-2}$, or $V^{N-4}$ potential. For the case of the partly filled atomic
shells we use the average over the non-relativistic configuration, as
described in \cite{BDT77}. Here we do not need this as both our potentials
correcpond to the closed shells $2p^6$, or $3s^2$. The effective Hamiltonian
has the form:
 \begin{equation}\label{eq2}
 H^\mathrm{eff}=H_{1}+H_{2}\,,
 \end{equation}
where $H_{1}$ represents the one-body part of the Hamiltonian, and $H_{2}$
represents the two-body residual Coulomb interaction.

In the CI+MBPT method we incorporate core excitations by including
perturbation theory terms into the effective Hamiltonian. The one-body part
$H_{1}$ is modified to include the correlation potential $\Sigma_{1}$, that
accounts for the core-valence correlations (it is also known as self-energy
correction):
 \begin{equation}\label{eq3}
 H_{1}\rightarrow H_{1}+\Sigma_{1}\,,
 \end{equation}
and the two-body term $H_{2}$ now includes the effective screening of the
two-body interactions by the core:
 \begin{equation}\label{eq4}
 H_{2}\rightarrow H_{2}+\Sigma_{2}\,.
 \end{equation}
Both $\Sigma_{1}$ and $\Sigma_{2}$ are calculated in the second-order MBPT
\cite{DFK96b}. In the same second order of MBPT there is also a three-electron
correction to the Hamiltonian $\Sigma_{3}$ \cite{DFK96b}, which describes
effective three-electron interactions between valence electrons induced by the
core polarization effects. This interaction is typically very small, but
becomes important for partly filled $d$, or $f$ shells \cite{BFK08,KSPT16}.

\begin{table*}
\caption{The energy levels (in cm$^{-1}$) obtained in $V^{N-2}$ and $V^{N-4}$
approximations are compared with the experiment and the results by Savukov
\cite{Sav15}. The energies are counted from the ground state.}
\label{tab:2}       
\begin{center}
\begin{tabular}[b]{ccrrrrrrr}
\hline \hline

    &   &   & \multicolumn{2}{c}{CI+MBPT}
    & &\multicolumn{2}{c}{Differences (\%)} \\
    \multicolumn{1}{c}{Conf.}
    &\multicolumn{1}{c}{Term}
    &\multicolumn{1}{c}{Exper.}
    &\multicolumn{1}{c}{$V^{N-2}$}
    &\multicolumn{1}{c}{$V^{N-4}$}
    & Ref.\ \cite{Sav15}
    &\multicolumn{1}{c}{$V^{N-2}$}
    &\multicolumn{1}{c}{$V^{N-4}$}
    & Ref.\ \cite{Sav15} \\
\hline
  3s$^{2}$3p$^{2}$ & $^{3}P_{1}$ & 77 & 76 & 75 & 80 &1.30 & 2.60 & -3.90 \\
  3s$^{2}$3p$^{2}$ & $^{3}P_{2}$ & 223 & 224 & 221 & 234 & -0.15 & 0.90 & -4.93 \\
  3s$^{2}$3p4s & $^{3}P^{o}_{0}$ & 39683 & 39773 & 39669 & 39201 & -0.23 & 0.04 & 1.21 \\
  3s$^{2}$3p4s & $^{3}P^{o}_{1}$ & 39760 & 39851 & 39747 & 39282 & -0.23  & 0.03 &1.20 \\
  3s$^{2}$3p4s & $^{3}P^{o}_{2}$ & 39955 & 40045 & 39941 & 39485 & -0.22 &0.04 & 1.18 \\
  3s$^{2}$3p4s & $^{1}P^{o}_{1}$ & 40992 & 41136 & 41003 & 40606 & -0.35 & -0.03 & 0.94 \\
  3s3p$^{3}$ & $^{3}D^{o}_{1}$ & 45276 & 45090 & 45144 & 44852 & 0.41  & 0.29  &0.94\\
  3s$^{2}$3p4p & $^{1}P_{1}$ & 47284 & 47519 & 47266  & 47611 & -0.50 & 0.04 &-0.69  \\
  3s$^{2}$3p4p & $^{3}D_{1}$ & 48020 & 47722 & 48020  & 48398 &0.62 & 0.00 & -0.79 \\
   \hline\hline
\end{tabular}%
\end{center}
\end{table*}

\section{RESULTS AND DISCUSSIONS}
\label{sec:1}

The basis set was constructed in the frame of the HFD approach. In the
$V^{N-2}$ approximation we start with solution of the Dirac-Fock equations for
the [1s$^{2}$2s$^{2}$2p$^{6}$3s$^{2}$] closed shells. As a next step, all
orbitals up $3s$ were frozen and 4-5$s$, 3-5$p$, 3-5$d$, 4-5$f$, and 5$g$
orbitals were constructed in the same $V^{N-2}$ potential. Higher virtual
orbitals were obtained with the help of B-splines \cite{SJ96}. The lower
components of the Dirac bispinors were formed using kinetic balance condition
(see, e.g.\ \cite{KPST15}). The MBPT corrections to the effective Hamiltonian
were calculated in the second order MBPT. Note that in this case we had to
include the so-called subtraction diagrams \cite{DFK96b}.

To illustrate the role of the core-valence correlations we calculated the
low-lying energy levels using pure valence CI in the frosen core approximation
and in the frame of the CI+MBPT method. The results of the energy level
calculations are presented in Table \ref{tab:1}. The energies are counted from
the ground state $[3s^23p^2]\,{}^3P_0$. The values are given in cm$^{-1}$. The
differences of our results with the experimental data from NIST \cite{NIST}
are given in the last columns to illustrate the accuracy of each approach. We
find that the accuracy of the CI method is mostly on the order of few percent,
while the accuracy of the CI+MBPT method is mostly better than 1\%. The lowest
relative accuracy is for the two singlet states ${}^1D_2$ and ${}^1S_0$ of the
configuration $3s^23p^2$. On the one hand, these levels lye rather close to
the ground state and the absolute accuracy for these levels is not so much
different from the others. On the other hand, for the singlet states the exact
wave function has a cusp at $r_{ik}=0$. To reproduce this cusp accurately one
needs very large CI space. Therefore, it is not surprising that we have not
saturated CI space for these states.

As discussed above, when we apply the CI+MBPT method to the atoms with more
than two valence electrons, the effective valence Hamiltonian includes
three-electron term $\Sigma_3$. Here we calculated respective corrections and
found them to be smaller than the overall theoretical error. We conclude that
three-electron interactions are insignificant for Si I. This is not surprising
as the core here does not include $d$, or $f$ shells and the valence $3d$
orbital is weakly bound and does not overlap with the core.

Recently the new accurate method to account for the QED corrections in the
polyvalent systems was suggested in Refs.\ \cite{STY15,TKSSD16}. We calculated
these corrections together with Breit corrections to the Coulomb interaction
and added them to our final results. In general these contributions appear to
be too small to influence our final accuracy. However, they slightly improve
theoretical fine structure.

We have also considered Si I as the four-valence atom using $V^{N-4}$
approximation. Calculations were quite similar, with the exception of the
construction of the basis set. The latter was formed by solving the Dirac-Fock
equations for the [1s$^{2}$2s$^{2}$2p$^{6}$] closed shells. Then, the orbitals
4-5$s$, 3-5$p$, 3-5$d$, 4-5$f$, and 5$g$ were constructed in the same
$V^{N-4}$ potential. For this potential the MBPT part did not include
subtraction diagrams as the potential $V^{N-4}$ corresponds to the bare core.
For CI calculations we used exactly the same sets of even- and odd-parity
configurations, as for the calculations in the $V^{N-2}$ approximation.

In Table \ref{tab:2} we compare results obtained within the CI+MBPT method in
the $V^{N-2}$ and $V^{N-4}$ approximations. For the neutral atom, the initial
$V^{N-4}$ approximation is clearly less accurate, than the $V^{N-2}$
approximation. For this reason the CI calculation here gives poorer agreement
with the experiment. However, Table \ref{tab:2} demonstrate that the accuracy
of the final CI+MBPT results are comparable. In fact, for the majority of
states the $V^{N-4}$ approximation is slightly better. Our results found here
are in good agreement with previous calculation by Savukov \cite{Sav15}, but,
in general, a little more accurate.

\section{CONCLUSION}

We have studied the accuracy of the CI+MBPT method for the four-valent Si. We
performed calculations in the $V^{N-2}$ and $V^{N-4}$ initial approximations
and studied the role of the Breit, QED, and three-electron corrections. Our
results show that both initial approximations lead to comparable final
accuracy, though the $V^{N-4}$ approximation is a little better. Both our
calculations have slightly higher accuracy than recent calculation in Ref.\
\cite{Sav15}. The QED and Breit corrections improve theoretical fine
structure, but do not improve the gross structure. The effective
three-electron interactions for Si are negligibly small. The overall accuracy
of the theory is on the level of 1\%. Before including higher order terms of
the MBPT it is necessary to saturate the CI space, which appears to be rather
costly for the four-valent systems.\\[3mm]


This work is partly supported by Russian Foundation for Basic Research Grant
No.\ 14-02-00241.


\begin{thebibliography}{22}

\bibitem{DFK96a}
V.A. Dzuba, V.V. Flambaum, M.G. Kozlov, JETP Lett. \textbf{63}, 882 (1996)

\bibitem{DFK96b}
V.A. Dzuba, V.V. Flambaum, M.G. Kozlov, Phys. Rev. A \textbf{54}, 3948 (1996)

\bibitem{PKRD01}
S.G. Porsev, M.G. Kozlov, Y.G. Rakhlina, A.~Derevianko, Phys. Rev. A
  \textbf{64}, 012508 (2001), \texttt{arXiv: physics/0102070}

\bibitem{SJ02}
I.M. Savukov, W.R. Johnson, Phys. Rev. A \textbf{65}, 042503 (2002)

\bibitem{PSK12}
S.G. Porsev, M.S. Safronova, M.G. Kozlov, Phys. Rev. Lett. \textbf{108},
173001
   \texttt{arXiv: 1201.5615}

\bibitem{KonKoz15}
E.A. Konovalova, M.G. Kozlov, Phys. Rev. A \textbf{92}, 042508 (2015),
  \texttt{1508.01958}

\bibitem{Dzu05a}
V.A. Dzuba, Phys. Rev. A \textbf{71}, 032512 (2005)

\bibitem{Dzu05b}
V.A. Dzuba, Phys. Rev. A \textbf{71}, 062501 (2005)

\bibitem{Ber11}
J.C. Berengut, Phys. Rev. A \textbf{84}, 052520 (2011), arXiv:1110.2292

\bibitem{SDFS14b}
M.S. Safronova, V.A. Dzuba, V.V. Flambaum, U.I. Safronova, S.G. Porsev, M.G.
  Kozlov, Phys. Rev. A \textbf{90}, 052509 (2014), \texttt{arxiv:1409.6124}

\bibitem{Sav16}
I.M. Savukov, Phys. Rev. A \textbf{93}, 022511 (2016)

\bibitem{PKST16}
S.G. Porsev, M.G. Kozlov, M.S. Safronova, I.I. Tupitsyn, Phys. Rev. A
  \textbf{93}, 012501 (2016), \texttt{1510.06679}

\bibitem{BFK08}
J.C. Berengut, V.V. Flambaum, M.G. Kozlov, J. Phys. B \textbf{41}, 235702
  (2008), \texttt{arXiv: 0806.3501}

\bibitem{KSPT16}
M.~Kozlov, M.~Safronova, S.~Porsev, I.~Tupitsyn, Phys. Rev. A \textbf{94},
  032512 (2016), \texttt{1607.05843}

\bibitem{NIST}
NIST, \emph{{Atomic Spectra Database}},
  \texttt{http://physics.nist.gov/PhysRefData/ASD/index.html}

\bibitem{BDT77}
V.F. Bratsev, G.B. Deyneka, I.I. Tupitsyn, Bull. Acad. Sci. USSR, Phys. Ser.
  \textbf{41}, 173 (1977)

\bibitem{KT87}
S.A. Kotochigova, I.I. Tupitsyn, J. Phys. B \textbf{20}, 4759 (1987)

\bibitem{KPST15}
M.~Kozlov, S.~Porsev, M.~Safronova, I.~Tupitsyn, Computer Physics
  Communications \textbf{195}, 199 (2015)

\bibitem{Sav15}
I.M. Savukov, Phys. Rev. A \textbf{91}, 022514 (2015)

\bibitem{SJ96}
J.~Sapirstein, W.R. Johnson, J. Phys. B \textbf{29}, 5213 (1996)

\bibitem{STY15}
V.~Shabaev, I.~Tupitsyn, V.~Yerokhin, Computer Physics Communications
  \textbf{189}, 175 (2015)

\bibitem{TKSSD16}
I.I. Tupitsyn, M.G. Kozlov, M.S. Safronova, V.M. Shabaev, V.A. Dzuba, arXiv
  (2016), \texttt{1607.07064}

\end{thebibliography}
%
%
%

\end{document}